\documentclass[aps,prl,superscriptaddress]{revtex4-1}

\usepackage{graphicx}
\usepackage{amsmath}
\usepackage{caption}
\usepackage{subcaption}
\usepackage{appendix}
\usepackage{float}

\begin{document} 
	
\title{ Effective Approximation of Electromagnetism for Axion Haloscope Searches}

\author{Younggeun Kim}
\affiliation{Center for Axion and Precision Physics Research, IBS, Daejeon, South Korea, 34051}
\affiliation{Department of Physics, KAIST, Daejeon, South Korea, 34141}

\author{Dongok Kim}
\affiliation{Center for Axion and Precision Physics Research, IBS, Daejeon, South Korea, 34051}
\affiliation{Department of Physics, KAIST, Daejeon, South Korea, 34141}

\author{Junu Jung}
\affiliation{Center for Axion and Precision Physics Research, IBS, Daejeon, South Korea, 34051}
\affiliation{Department of Physics, KAIST, Daejeon, South Korea, 34141}

\author{Jinsu Kim}
\affiliation{Center for Axion and Precision Physics Research, IBS, Daejeon, South Korea, 34051}
\affiliation{Department of Physics, KAIST, Daejeon, South Korea, 34141}

\author{Yun Chang Shin\thanks{corresponding author}}
\email[]{corresponding author : yunshin@ibs.re.kr}
\affiliation{Center for Axion and Precision Physics Research, IBS, Daejeon, South Korea, 34051}

\author{Yannis K. Semertzidis}
\affiliation{Center for Axion and Precision Physics Research, IBS, Daejeon, South Korea, 34051}
\affiliation{Department of Physics, KAIST, Daejeon, South Korea, 34141}
 
\date{\today}

\begin{abstract}
We applied an effective approximation into Maxwell's equations including axion-photon interaction for haloscope searches. A set of Maxwell's equations acquired from this approximation exactly describes the reacted fields generated from the axion-photon interaction. Unlike other approaches, this set of Maxwell's equations inherently satisfies the boundary conditions for haloscope searches. Electromagnetic fields in cylindrical and toroidal cavities were evaluated from the Maxwell's equations including when the axion mass becomes ultra-light (sub-meV). Stored energy in both cavities was also examined.  A small but non-zero difference between the electric and magnetic stored energies appeared in both cases. The difference may come from  non-dissipating current induced by oscillating axions. 
\end{abstract}

\maketitle

\section{\label{sec:level1}Introduction}
An axion is a hypothetical particle suggested by the Peccei-Quinn (PQ) mechanism as a solution of the strong CP problem in the Standard Model\cite{1977PhRvL..38.1440P}. Phenomenological searches indicate that axions could be invisible because of their weak coupling with matter\cite{2016PhR...643....1M,1983PhLB..120..127P}. If axions are indeed invisible, they could play an important role in the composition of dark matter and be ubiquitous in our Universe. This invisibility of the axion has been described by two different axion models, the KSVZ and DFSZ\cite{1979PhRvL..43..103K,1980NuPhB.166..493S,1981PhLB..104..199D,Zhitnitsky:1980tq}.

In addition to the very weak coupling of axions with particles from the Standard Model, they may be weakly interacting with electromagnetic fields as well\cite{1985NuPhB.260..689S,1951PhRv...81..899P}. The interaction of an axion with an electromagnetic field is governed by allowing the coupling of the axion to the electromagnetic field as
\begin{equation}
 a\mathbf{E}\cdot\mathbf{B} \propto-a F_{\mu\nu}\tilde{F}^{\mu\nu}.
 \label{eq1-1}
\end{equation}
This coupling results in a conversion of axions into photons via the inverse Primakoff effect\cite{1951PhRv...81..899P}. Most of the successful experiments searching for axion are based on this axion-photon interaction in addition to an assumption of axions as halo dark matter, which are accordingly called axion haloscope searches\cite{1983PhRvL..51.1415S, 1987PhRvL..59..759V}. Classical Maxwell's equations need to be modified to include the interaction of axion with electromagnetic fields\cite{1987PhRvL..58.1799W,2013MPLA...2850162V,2015MPLA...3050204T}. 
 
Due to the axion anomaly, however, this modification of Maxwell's equations causes the other issue: it doesn't naturally satisfy certain boundary condition, particularly one necessary for axion haloscope searches\cite{2016PhRvL.116p1804M,2016PhRvD..94k1702K}. This is mainly because an electromagnetic field generated from the axion interaction is not clearly separated from applied external fields, which is necessary to create the haloscope condition, in Maxwell's equations. 

In this paper, we introduce an effective approximation of Maxwell's equations that can decouple the reacted electromagnetic field generated by the axion interaction from the external fields. The separated Maxwell's equations provide the motion of the reacted electromagnetic field only. We applied them to axion haloscope cases, and showed they both naturally satisfy boundary conditions for haloscope searches. The reacted electromagnetic field in a cylindrical cavity as well as a toroidal cavity were evaluated from the separated Maxwell's equations. Electric and magnetic energies stored in cavity modes were also estimated from this approximation. A very small but non-zero value arises in the difference between the electric and magnetic stored energies in both cases. The difference can be interpreted as a polarization density induced by oscillating axion.

\section{\label{sec:level1}Separation of Maxwell's Equations for Haloscope Searches}
The effective Lagrangian describing the axion electromagnetic interaction including an axion-like term can be derived in SI units as
\begin{equation}
\mathcal{L}_{0+a}=-\frac{1}{4\mu_0}\mathit{F}^{\mu\nu}\mathit{F}_{\mu\nu}+\frac{g_{a\gamma \gamma }}{4\mu_0}a\mathit{F}^{\mu\nu}\tilde{F}_{\mu\nu}-\mathit{A}_\mu\mathit{J}^{\mu}_e+\mathcal{L}_U,
\label{eq2-2}
\end{equation}
where $\mathcal{L}_{0}=-\frac{1}{4\mu_0}\mathit{F}^{\mu\nu}\mathit{F}_{\mu\nu}-\mathit{A}_\mu\mathit{J}^{\mu}_e$ is the classical EM Lagrangian and $\mathcal{L}_{U}=\frac{1}{2}(\partial^{\mu}a)(\partial_{\mu}a)-U(a)=\frac{1}{2}(\partial^{\mu}a)(\partial_{\mu}a)-\frac{1}{2}\omega_{a}^2a^2$ is the axion Lagrangian with a potential $U(a)$.

The axion-like term in the Lagrangian is 
\begin{equation}
{\mathcal{L}_a} = \frac{{{g_{a\gamma \gamma }}}}{{4{\mu _0}}}a{F^{\mu \nu }}{\tilde{F}_{\mu \nu }} = - \frac{{{g_{a\gamma \gamma }}}}{{{\mu _0 c}}}a\mathbf{E} \cdot \mathbf{B},
\label{eq2-3}
\end{equation}
where $a$ is axion field. $g_{a\gamma \gamma}$ is the two-photon coupling to axion field in unit of $\textrm{GeV}^{-1}$ defined as
\begin{equation}
g_{a\gamma \gamma}=\frac{\alpha_{EM}}{2\pi f_{a}}c_{a\gamma\gamma},
\label{eq2-4}
\end{equation}
where $\alpha_{EM}\approx1/137$ is the fine structure constant and $f_{a}$ is the axion decay constant in unit of $\mathrm{GeV}$\cite{2016PhR...643....1M}. $c_{a\gamma\gamma}$ is the dimensionless coupling which is model dependent as $c_{a\gamma\gamma}=-1.92 ~(\textrm{KSVZ})$, or $c_{a\gamma\gamma}=0.75 ~(\textrm{DFSZ})$\cite{1979PhRvL..43..103K,1980NuPhB.166..493S,1981PhLB..104..199D,Zhitnitsky:1980tq}.  For QCD axion, the $g_{a\gamma \gamma}$ is defined by the breaking scale  $f_{a}$ of the Peccei-Quinn (PQ) symmetry. For generic axion like particle (ALP), the coupling $g_{a\gamma \gamma}$ is treated as free parameter which is assumed to be much smaller than one for the QCD axion in certain ALP models\cite{2014JHEP...06..037D}.

In both cases, Maxwell's equations for the axion-photon interaction can be derived from the full Lagrangian written in Eq.\ref{eq2-3} with the Bianchi Identity as follows:
\begin{subequations}
\label{eq2-7}
\begin{align}
&\nabla \cdot \left({\mathbf{E}-c g_{a\gamma \gamma} a\mathbf{B}}\right)=\frac{\rho_{e}}{\varepsilon}, \label{eq2-7-a}\\ 
&\nabla \cdot \mathbf{B}=0, \label{eq2-7-b}\\
&\nabla\times\mathbf{E}=- \frac{\partial\mathbf{B}}{\partial t} , \label{eq2-7-c}\\
&\nabla\times \left({c\mathbf{B}+g_{a\gamma \gamma}a\mathbf{E}}\right)=\frac{1}{c}\frac{\partial}{\partial t} \left({\mathbf{E}-cg_{a\gamma \gamma} a\mathbf{B}}\right)+c\mu\mathbf{J}_{e}\label{eq2-7-d}.
\end{align}
\end{subequations}
Many of the ongoing experimental searches for axions from inverse Primakoff's effect are all based on the Maxwell's equations in Eq.\ref{eq2-7}\cite{,1951PhRv...81..899P,1983PhRvL..51.1415S}. However, because of the axion anomaly, this set of Maxwell's equations doesn't naturally satisfy certain boundary conditions. 

The general haloscope conditions that have been assumed in many previous approaches are\cite{2016PhRvL.116p1804M, 2016PhRvD..94k1702K} 
\begin{itemize}
\item zero current : $\mathbf{J_{e}}=0$,
\item zero charge density : $\rho_{e}=0$,
\item zero external electric field : $\mathbf{E}_{\textrm{ext}}=0$,
\item nonzero external magnetic field : $\mathbf{B}=\mathbf{B}_{\textrm{ext}}$,
\item curlless magnetic field : $\nabla\times\mathbf{B}=\nabla\times\mathbf{B}_{\textrm{ext}}=0$, 
\item time independent magnetic field : $\dot{\mathbf{B}}=0$.
\end{itemize}

Under these haloscope conditions, th Eq.\ref{eq2-7-d} becomes 
\begin{equation}
\nabla\times \mathbf{B}=-\frac{g_{a\gamma \gamma}}{c}\mathbf{B}_{\textrm{ext}}\frac{\partial \mathbf{ a}}{\partial t}.
\label{eq2-14}
\end{equation}
While the boundary condition constrains $\nabla\times \mathbf{B}=0$ on the left side of Eq.\ref{eq2-14}, the right side doesn't intuitively become zero because of the axion interaction term. One can't naively exclude the anomaly term either. 

In Refs.\cite{2016PhRvL.116p1804M, 2016PhRvD..94k1702K}, this problem was avoided by forcing the non-zero term in Eq.\ref{eq2-14} as a relationship of new fields $\mathbf{E}_{a}$ and $\mathbf{B}_{a}$ as
\begin{equation}
 \nabla \times\mathbf{B}_{a}=-\frac{g_{a\gamma\gamma}}{c}\mathbf{B}_{\textrm{ext}}\frac{\partial a}{\partial t}=\frac{1}{c^{2}}\frac{\partial\mathbf{E}_{a}}{\partial t}, 
 \label{eq2-7-1-1}
 \end{equation}
 where $\mathbf{E}_{a}$ and $\mathbf{B}_{a}$ are the electric and magnetic field components of the photons produced via the axion-photon interaction.  At the same time, the other relationship between $\mathbf{E}_{a}$ and $\mathbf{B}_{a} $ was also defined from Eq.\ref{eq2-7-c} with boundary conditions as
 \begin{equation}
\nabla \times\mathbf{E}_{a}=-\frac{\partial\mathbf{B}_{a}}{\partial t}.
\label{eq2-7-1-2}
\end{equation}

Eq.\ref{eq2-7-1-2} is Maxwell-Faraday equation, which is a generalization of Faraday's law. Dynamical electromagnetic fields defined in any space from given boundary conditions have to satisfy this relationship. In Refs.\cite{2016PhRvL.116p1804M, 2016PhRvD..94k1702K}, the $\mathbf{E}_{a}$ was estimated  from the right side of the relationship in Eq.\ref{eq2-7-1-1} as 
\begin{equation*}
-\frac{g_{a\gamma\gamma}}{c}\mathbf{B}_{\textrm{ext}}\frac{\partial a}{\partial t}=\frac{1}{c^{2}}\frac{\partial\mathbf{E}_{a}}{\partial t}.
\end{equation*}

$\mathbf{E}_{a}$ obtained from the relationship is
\begin{equation}
\mathbf{E}_{a}=-cg_{a\gamma\gamma} a\mathbf{B}_{\textrm{ext}}+\mathbf{f}(\mathbf{r}), 
\label{eq2-7-5}
\end{equation}
where the function $\mathbf{f}(\mathbf{r})$ can have only position dependence. At the same time, $\mathbf{B}_{a}$ in Refs.\cite{2016PhRvL.116p1804M, 2016PhRvD..94k1702K} was estimated by applying Stoke's theorem in the left side of relationship in Eq.\ref{eq2-7-1-1} as follows:
\begin{equation}
\int_{\cal{A}}(\nabla \times\mathbf{B}_{a})\cdot d\mathbf{A}=\oint_{\partial\cal{A}}\mathbf{B}_{a}\cdot d\mathbf{l}=-\frac{g_{a\gamma\gamma}}{c}\dot{ a}{\cal{A}}\mathbf{B}_{\textrm{ext}} ,
\label{eq2-7-2}
\end{equation}
where $\cal{A}$ is the area to be integrated, and we assume an uniform external magnetic field $\mathbf{B}_{\textrm{ext}}=\textrm{B}_{0}\hat{z}$. If we consider a certain geometry which has a rotational symmetry along the z axis such as a cylindrical cavity, $\mathbf{B}_{a}$ can be estimated from Eq.\ref{eq2-7-2} as
\begin{equation}
\mathbf{B}_{a}=-\frac{g_{a\gamma\gamma}}{2c}r{\textrm{B}_{0}}\dot{ a}\hat{\phi}.
\label{eq2-7-3}
\end{equation}
If the $\mathbf{E}_{a}$ and $\mathbf{B}_{a}$ are complete solutions, they should satisfy the Maxwell-Faraday's law in Eq.\ref{eq2-7-1-2}. To check it, having Eqs. \ref{eq2-7-5} and \ref{eq2-7-3} in Eq.\ref{eq2-7-1-2} becomes
\begin{equation}
\nabla \times\mathbf{E}_{a}=\nabla \times\left(-cg_{a\gamma\gamma} a\mathbf{B}_{0}+\mathbf{f}(\mathbf{r})\right)=-\frac{g_{a\gamma\gamma}}{2c}r\textrm{B}_{0}\ddot{ a}\hat{\phi}.
\label{eq2-7-4}
\end{equation}
Since we assume that $\mathbf{f}$ is a function of only $\mathbf{r}$, no solution for $\mathbf{f}(\mathbf{r})$ is possible to satisfy Eq.\ref{eq2-7-4} unless $\ddot{ a}=0$. Otherwise, these solutions for an electromagnetic field obtained from Eq.\ref{eq2-7-1-1} don't explicitly satisfy Maxwell-Faraday's law. A detailed check of Maxwell-Faraday's law can be found in the Appendix. 

As we have shown, Eq.\ref{eq2-7-1-1} posecessess a fundamentally incomplete relationship. Therefore, one can't evaluate electromagnetic fields directly from both relationships in Eq.\ref{eq2-7-1-1}. This is because the electromagnetic field generated from the axion-photon interaction was not clearly decoupled from the external electromagnetic field applied for the axion interaction in Maxwell's equations. 

This issue can be resolved by applying an effective approximation into the Maxwell's equations in Eq.\ref{eq2-7} to decouple them\cite{2018APS..APRY09006K, 14thpatras2018}. By assuming that the axion anomaly slightly perturbs the electromagnetic field, one can apply following relationships onto the electromagnetic field as
\begin{subequations}
\label{eq2-8}
\begin{align}
&\mathbf{E}=\sum_{m}(g_{a\gamma\gamma})^m\mathbf{E}_m=\mathbf{E}_{{0}}+g_{a\gamma\gamma}\mathbf{E}_{1} + g_{a\gamma\gamma}^{2}\mathbf{E}_{2} +\cdot \cdot ,\\
&\mathbf{B}=\sum_{m}(g_{a\gamma\gamma})^m\mathbf{B}_m=\mathbf{B}_{{0}}+g_{a\gamma\gamma}\mathbf{B}_{1} + g_{a\gamma\gamma}^{2}\mathbf{B}_{2} +\cdot \cdot .
\end{align}
\end{subequations}

Since $|g_{a\gamma\gamma}|$ is roughly $\sim10^{-8}\mathrm{GeV^{-1}}$ or less for both QCD axion and ALP cases, the higher order terms are ignored and only the leading and first order terms can be considered as
\begin{subequations}
\begin{align}
&\mathbf{E}\simeq\mathbf{E}_{{0}}+g_{a\gamma\gamma}\mathbf{E}_{1} ,\\
&\mathbf{B}\simeq\mathbf{B}_{{0}}+g_{a\gamma\gamma}\mathbf{B}_{1} .
\end{align}
\label{eq2-8-1}
\end{subequations}
If we input field relations in Eq.\ref{eq2-8-1} into the Maxwell's equations in Eq.\ref{eq2-7}, and ignores the ${g_{a\gamma\gamma}}^{2}$ terms again, the set of equations is reduced as
\begin{subequations}
\label{eq2-10}
\begin{align}
&\nabla \cdot \left( \mathbf{E}_{0}+g_{a\gamma\gamma}\mathbf{E}_{1}\right) = \rho_{e}/\epsilon_{0},\\
&\nabla \cdot \left( \mathbf{B}_{0}+g_{a\gamma\gamma}\mathbf{B}_{1}\right) = 0,\\
&\nabla \times \left(\mathbf{E}_{0}+g_{a\gamma\gamma}\mathbf{E}_{1}\right) =-\frac{\partial}{\partial t} \left(\mathbf{B}_{0}+g_{a\gamma\gamma}\mathbf{B}_{1} \right),\\
&\nabla\times (\mathbf{B}_{0}+g_{a\gamma\gamma}\mathbf{B}_{1}+ \frac{1}{c} g_{a\gamma\gamma} a \mathbf{E}_{{0}}) =\nonumber\\
&\frac{1}{c^{2}}\frac{\partial}{\partial t}(\mathbf{E}_{0}+g_{a\gamma\gamma}\mathbf{E}_{1} - c g_{a\gamma\gamma} a \mathbf{B}_{{0}}) +\mu_{0}\mathbf{J}_{e}.
\end{align}
\end{subequations}

Now one can assume a certain space filled with the external electromagnetic field, $\mathbf{E}_{0}$ and $\mathbf{B}_{0}$. If there is no axion, there is no axion-photon interaction. Therefore, the total electromagnetic field in the space will still remain $\mathbf{E}=\mathbf{E}_{0}$ and $\mathbf{B}=\mathbf{B}_{0}$. In this case, none of the terms in $g_{a\gamma\gamma}$ generated from the axion interaction in Eq.\ref{eq2-10} survive. Then, Eq.\ref{eq2-10} will be reduced into Maxwell's equations describing $\mathbf{E}_{{0}}$ and $\mathbf{B}_{{0}}$ only.

However, if there are interactions between the external electromagnetic field and axions through the axion-photon interaction term, a reacted electromagnetic field, $\mathbf{E}_{\mathrm{rea}}$, $\mathbf{B}_{\mathrm{rea}}$, will be generated. Therefore, the total electric and magnetic fields $\mathbf{E}$ and $\mathbf{B}$ in the space are slightly different from the external applied fields $\mathbf{E}_{0}$ and $\mathbf{B}_{0}$. In this case, the total fields can be expressed as a superposition of the original fields $\mathbf{E}_{0}$, $\mathbf{B}_{0}$ and the reacted fields from the axion-interaction term, $\mathbf{E}_{\mathrm{rea}}$, $\mathbf{B}_{\mathrm{rea}}$ as
\begin{subequations}
\begin{align}
&\mathbf{E}=\mathbf{E}_{0}+\mathbf{E}_{\mathrm{rea}} ,\\
&\mathbf{B}=\mathbf{B}_{0}+\mathbf{B}_{\mathrm{rea}} .
\end{align}
\label{eq2-11}
\end{subequations}
By recalling Eq.\ref{eq2-8-1}, one can set the relationships of the reacted fields as
\begin{subequations}
\begin{align}
\mathbf{E}_{\mathrm{rea}}&=g_{a\gamma\gamma}\mathbf{E}_{1},\\
\mathbf{B}_{\mathrm{rea}}&=g_{a\gamma\gamma}\mathbf{B}_{1}.
\end{align}
\end{subequations}

Now, we can apply the specific boundary condition for axion haloscope case: a curlless time independent external magnetic field ($\nabla\times\mathbf{B}_{0}=0,\ \dot{\mathbf{B}}_{0}=0$), zero external electric field ($\mathbf{E}_{0}=0$) with zero current, ($\mathbf{J_{e}}=0$), and zero charge density, ($\rho_{e}=0$). The set of Maxwell's equations describing the external electromagnetic field, $\mathbf{E}_{0}$ and $\mathbf{B}_{0}$, remains as
\begin{equation*}
\begin{aligned}
\nabla \cdot \mathbf{B}_{0} = 0,\\
\nabla\times \mathbf{B}_{0} = 0,
\end{aligned}
\label{eq2-12}
\end{equation*}
which still satisfy the original boundary condition for the haloscope experiment. 

Furthermore, a new set of Maxwell's equations which describes only reacted fields, $\mathbf{E}_{\mathrm{rea}}$, $\mathbf{B}_{\mathrm{rea}}$ to the applied external magnetic field $\mathbf{B}_{0}$ can be obtained as
\begin{subequations}
\begin{align}
&\nabla \cdot \mathbf{E}_{\mathrm{rea}} = 0 ,\\
&\nabla \cdot \mathbf{B}_{\mathrm{rea}} = 0 ,\\
&\nabla \times \mathbf{E}_{\mathrm{rea}} =-\frac{\partial \mathbf{B}_{\mathrm{rea}}} {\partial t} , \\
&\nabla\times \mathbf{B}_{\mathrm{rea}} = \frac{1}{c^{2}} \frac{\partial}{\partial t} (\mathbf{E}_{\mathrm{rea}} -c g_{a\gamma\gamma} a \mathbf{B}_{0}) .
\end{align}
\label{eq2-13}
\end{subequations}
Eq.\ref{eq2-13} is different from Maxwell's equations in Eq.\ref{eq2-7}. First of all, Eq.\ref{eq2-13} is obtained from a perturbation of the total electromagnetic field. Second, while it is not possible to isolate the reacted electromagnetic field from the external electromagnetic field in Eq.\ref{eq2-7}, they are clearly decoupled in Eq.\ref{eq2-13}. In the following sections, we will present the solution of the Maxwell's equations in Eq.\ref{eq2-13} in two different resonant cavity cases.

\section{\label{sec:level1}Electromagnetic field for Haloscope Searches}
Since the axion field in the haloscope experiment can be considered as homogeneous one, $ a$ can be expressed as $a(t)=a_{0}e^{-i{\omega_{a}}t}$ where $\omega_{a}$ is the oscillating frequency of the axion\cite{1983PhRvL..51.1415S, 2016PhR...643....1M}. Unless one takes a decoupling limit of the axion where $\partial_{t}a (t) \rightarrow 0$, this assumption is still valid for any axion model with non-zero axion mass constrained from cosmological observations\cite{2015PhRvD..91k1702H,2016PhRvD..93b5007H}. From the local galactic dark matter density of $\sim 0.3\ \textrm{GeV}/ \textrm{cm}^{3}$, $\frac{a_{0}}{f_{a}}\approx 3.7\times10^{-19}$ was assumed\cite{1986PhRvD..33..889T}. In addition, $\mathbf{E}_{\mathrm{rea}}$ and $\mathbf{B}_{\mathrm{rea}}$ should oscillate with the form of $e^{-i\omega_at}$ so that $\mathbf{E}_{\mathrm{rea}}=\mathbf{E}_{\mathrm{rea}}(r, t)=\mathbf{E}_{\mathrm{rea}}(r)e^{-i\omega_at}$, $\mathbf{B}_{\mathrm{rea}}=\mathbf{B}_{\mathrm{rea}}(r, t)=\mathbf{B}_{\mathrm{rea}}(r)e^{-i\omega_at}$. 
\subsection{\label{sec:citeref}Cylindrical cavity}
The most common geometry of the resonant cavity for the axion haloscope search is a cylindrical shape as in ADMX or CULTASK\cite{2010PhRvL.104d1301A,2017arXiv171001833S,2017PhRvD..95f3017O,2016chep.confE.197W}. The boundary condition for a cylindrical resonant cavity with radius $r_{0}$ requires a uniformly applied magnetic field ($\mathbf{B}_{0}=\textrm{B}_0\hat{z}$) and a conductive surface for the resonant cavity. Here, we assume perfect electric conductor (PEC) for the cavity surface. For simplicity, we assumed the uniform external magnetic field $\mathbf{B}_{0}$ is applied only inside the cavity which means $\mathbf{B}_{\mathrm{ext}}=\mathbf{B}_{0}$ for $ r\leq r_{0}$ and $\mathbf{B}_{\mathrm{ext}}=0$ for $ r>r_{0}$. 

The wave equation for reacted fields can be obtained by applying curl for Eq.\ref{eq2-13} as,
\begin{subequations}
\begin{align}
&{{ \nabla ^{2} \mathbf{E}_{\mathrm{rea}}=\frac{\partial^{2}}{c^{2}\partial t^{2}}\mathbf{E}_{\mathrm{rea}}-\frac{g_{a\gamma\gamma}}{c}\mathbf{B}_{0}\frac{\partial^{2}}{\partial t^{2}}a(t)}},\\
&{{ \nabla ^{2} \mathbf{B}_{\mathrm{rea}}=\frac{\partial^{2}}{c^{2}\partial t^{2}}\mathbf{B}_{\mathrm{rea}}}}.
\end{align}
\label{eq3-1}
\end{subequations}
From Eq.\ref {eq3-1}, a special solution for the electric field can be defined as $\mathbf{E}_{\mathrm{rea}}^{\mathrm{special}}=cg_{a\gamma\gamma} a_{0}\mathbf{B}_{0}e^{-i\omega_{a}t}$. Combining with ordinary solutions, the full expression of electric and magnetic fields becomes
\begin{subequations}
\begin{align}
&\mathbf{E}_{\mathrm{rea}}=\left(c g_{a\gamma\gamma} a_{0}\textrm{B}_{0}e^{-i\omega_{a}t}+Ce^{-i\omega_{a}t}\mathbf{J}_{0}(kr)\right)\hat{z},\\
&\mathbf{B}_{\mathrm{rea}}=-iC\frac{k}{\omega_{a}} e^{-i\omega_{a}t}\mathbf{J}_{1}(kr)\hat{\phi},
\end{align}
\label{eq3-3}
\end{subequations}
where $k$ satisfies $k=\frac{\omega_{a}}{c}$. The coefficient $C$ is determined from the boundary condition for the conductive cavity in which the tangential component of the electric field must be terminated at the surface of the conductive cavity, $\mathbf{E}_{\mathrm{rea}}^{||}(r_{0})=0$. Then, the electric and magnetic fields inside the cavity $(r<r_{0})$ become
\begin{subequations}
\begin{align}
&\mathbf{E}_{\mathrm{rea}}=c g_{a\gamma\gamma} a_{0}\textrm{B}_{0}e^{-i\omega_{a}t}\left(1-\frac{\mathbf{J}_{0}(kr)}{\mathbf{J}_{0}(kr_{0})}\right)\hat{z},\\
&\mathbf{B}_{\mathrm{rea}}=ig_{a\gamma\gamma} a_{0}\textrm{B}_{0}e^{-i\omega_{a}t}\left(\frac{\mathbf{J}_{1}(kr)}{\mathbf{J}_{0}(kr_{0})}\right)\hat{\phi}.
\end{align}
\label{eq3-4}
\end{subequations}
This set of solutions is different not only from the ordinary resonant solutions but also from the solutions in Ref.\cite{2016PhRvL.116p1804M}. As shown in Eq.\ref{eq3-3}, the special solution for the reacted electric field, $\mathbf{E}_{\mathrm{rea}}^{\mathrm{special}}=c g_{a\gamma\gamma} a_{0}\textrm{B}_{0}e^{-i\omega_{a}t}$, doesn't propagate or resonate by itself. It just oscillates with the axion oscillating frequency, $\omega_{a}$. 

The electromagnetic field in Eq.\ref{eq3-3} starts to propagate or resonate only when the ordinary solutions are acquired. This implies that a conductive surface is required as a boundary condition to define the reacted electromagnetic field since the amplitude of reacted fields is defined only by the boundary condition. Same results with Eq.\ref{eq3-4} were obtained in Ref.\cite{2015PhRvD..91k1702H,2016PhRvD..93b5007H} although the result was interpreted as different effect.

For ultralight (sub-$\mathrm{meV}$) axions or ALPs, the Compton wavelength of the axion is much longer than the size of the resonant cavity so that one can take $kr\ll 1$. In this limit, Eq.\ref{eq3-4} can be approximated as
\begin{subequations}
\begin{align}
&\mathbf{E}_{\mathrm{rea}}\approx cg_{a\gamma \gamma} a \mathrm{B}_{0}(\frac{kr}{2})^{2}\hat{z},\ (r<r_{0})\\
&\mathbf{B}_{\mathrm{rea}}\approx  ig_{a\gamma\gamma} a_{0}\mathrm{B}_{0}e^{-i\omega_{a}t}\frac{kr}{2}\hat{\phi}=-\frac{g_{a\gamma\gamma}}{2c}r{\textrm{B}_{0}}\dot{ a}\hat{\phi}\ (r<r_{0}).
\end{align}
\label{eq3-4-1}
\end{subequations}
The magnetic field in Eq.\ref{eq3-4-1} is same as the result from Ref.\cite{2014PhRvL.112m1301S}.

\subsection{\label{sec:citeref}Toroidal cavity}
The toroidal cavity has been considered a very attractive geometry for axion haloscope searches because the conversion power would be improved by having the rather large volume of toroidal geometry. Recently, there was a proposed experiment looking for an axion with a prototype toroidal resonant cavity called ACTION at CAPP\cite{2017PhRvD..96f1102C}.

We consider a toroidal cavity having a geometry with the major radius $R$ and minor radius $r_{0}$ as a resonant cavity. The simple toroidal coordinate is $\left(r,\vartheta,\varphi\right)$ where $\vartheta$ is the counter-clockwise angle in the vertical section of the toroid, and $\varphi$ is the azimuthal angle. The transformation to the ordinary Cartesian coordinate will be

\begin{equation*}
\begin{aligned}
&x=\left(R+r\cos\vartheta\right)\cos\varphi,\\
&y=\left(R+r\cos\vartheta\right)\sin\varphi,\\
&z=r\sin\vartheta.
\end{aligned}
\end{equation*}
For the toroidal cavity, the external magnetic field, $\mathbf{B}_{\mathrm{ext}}$, is applied along the $\hat{\varphi}$ direction as
\begin{equation*}
\mathbf{B}_{\mathrm{ext}}=\frac{\textrm{B}_{0}}{R+r\textrm{cos}\vartheta}\hat{\varphi},
\end{equation*}
which still satisfies the boundary condition for the haloscope searches, $\nabla\times \mathbf{B}_{\mathrm{ext}} = 0,\ \dot{\mathbf{B}}_{\mathrm{ext}}=0$, in a toroidal cavity. 

Since the differential equation in Eq.\ref{eq3-1} is not separable in the toroidal coordinate, we applied a trial solution in  approximate form. The inseparable solution $W$ for wave equation has a dependence of $r$, and $\vartheta$ as $W(r,\vartheta)$. The differential equation that the $W(r,\vartheta)$ must satisfy is
\begin{equation}
\begin{aligned}
\Bigg[\frac{\partial^{2}}{\partial r^{2}}&+\frac{\partial^{2}}{r^{2}\partial \vartheta^{2}}+\left(\frac{1}{r}+\frac{\cos\vartheta}{R+r\cos\vartheta}\right)\frac{\partial}{\partial r}\\
&-\frac{\sin\vartheta}{r(R+r\cos\vartheta)}\frac{\partial}{\partial\vartheta}+k^{2}\Bigg]W(r,\vartheta)=0.\\
\end{aligned}
\label{eq3-5}
\end{equation}

As we assume leading order of $W(r,\vartheta)$ as $W(r,\vartheta)\approx\frac{f(r)}{h}=\frac{f(r)}{R+r\cos\vartheta}$, a set of solutions for electric and magnetic fields can be acquired as
\begin{subequations}
\begin{align}
&\mathbf{E}_{\mathrm{rea}}=cg_{a\gamma\gamma} a_{0} e^{-i\omega_{a} t}\left({\frac{\textrm{B}_{0}}{R+r\cos\vartheta}}\right)\left(1-\frac{\mathbf{J}_{0}(kr)}{\mathbf{J}_{0}(kr_{0})}\right)\hat{\varphi},\label{eq3-7-a}\\
&\mathbf{B}_{\mathrm{rea}}=i g_{a\gamma\gamma} a_{0} e^{-i\omega_{a} t}\left({\frac{\textrm{B}_{0}}{R+r\cos\vartheta}}\right)\left(\frac{\mathbf{J}_{1}(kr)}{\mathbf{J}_{0}(kr_{0})}\right)\hat{\vartheta}.\label{eq3-7-b}
\end{align}
\label{eq3-7}
\end{subequations}
The solution for electromagnetic field is almost identical with one for the cylindrical case. But, due to the approximation we took for the trial solution in the toroidal case, this set of solutions for electric and magnetic fields in Eq.\ref{eq3-7} has certain limitation in terms of accuracy. Especially when the inverse aspect ratio ($\eta = r_{0}/R$) is larger than 0.5, solutions from this approximation become inappropriate. However, since it is not realistic to have a toroidal cavity with the inverse aspect ratio larger than 0.5, this approximation is still valid for a practical purpose. In $kr\ll 1$ limit, the electromagnetic fields have very similar form with Eq.\ref{eq3-4-1}.

For a  toroidal geometry with rectangular cross section\cite{2016PhRvL.117n1801K} with inner radius $r_{1}$, outer radius $r_{2}$, and height $h$, one can consider two different cases, PEC toroidal cavity or non PEC toroidal cavity.   For simplicity, we also assume that the external magnetic field is provided only between two cavity surfaces, $r_{1}$ and $r_{2}$ which means $\mathbf{B}_{\mathrm{ext}}=\mathrm{B}_{0}\frac{r_{1}}{r}\hat{\varphi}$ for $ r_{1}\leq r\leq r_{2}$ and $\mathbf{B}_{\mathrm{ext}}=0$ for  $r>r_{2}$ or $0\leq r< r_{1}$. 

For PEC cavity case, the electric field in the region of interest ($r_{1}\leq r \leq r_{2}$)  has following solution,
\begin{equation}
\mathbf{E}_{\mathrm{rea}}=\left(\frac{cg_{a\gamma \gamma} a \mathrm{B}_{0}r_{1}}{r}+A\mathbf{J}_{1}(kr)+B\mathbf{Y}_{1}(kr)\right)\hat{\varphi}.
\end{equation}
Since we assume PEC for the both ends of cavity surfaces at $r_{1}$ and $r_{2}$, one can use $\mathbf{E}_{\mathrm{rea}}(r_{1})=\mathbf{E}_{\mathrm{rea}}(r_{2})=0$ to determine unknown coefficients, A and B. For $kr\ll 1$ limit, we can get them as $A=0$, $ B=\frac{1}{2} {\pi kcg_{a\gamma \gamma} a \mathrm{B}_{0}r_{1}}$, respectively. 

The corresponding magnetic field  in this region becomes,
\begin{equation}
\mathbf{B}_{\mathrm{rea}}=\frac{g_{a\gamma \gamma}\dot{ a}\mathrm{B}_{0}}{c}r_{1}\left(\ln (\frac{kr}{2})+\gamma\right)\hat{z},\ (r_{1}<r<r_{2}).
\end{equation}

In the central region of the rectangular toroidal cavity where $0\leq r<r_{1}$, the electric field has a solution of $A\mathbf{J}_{1}(kr)$ and for $kr\ll 1$ limit, the electromagnetic field approximately becomes,
\begin{subequations}
\begin{align}
&\mathbf{E}_{\mathrm{rea}}=A\mathbf{J}_{1}(kr)\hat{\varphi}\simeq\frac{A}{2}kr \hat{\varphi}, \ (0\leq r<r_{1}),\label{eq3-8-1-a}\\
&\mathbf{B}_{\mathrm{rea}}=\frac{A}{ic}\mathbf{J}_{0}(kr)\hat{z}\simeq\frac{A}{ic}\hat{z}, \ (0\leq r<r_{1}) \label{eq3-8-1-b}.
\end{align}
\label{eq3-8-1}
\end{subequations}
The electric field in Eq.\ref{eq3-8-1-a} should be terminated at $r=r_{1}$ to satisfy the PEC boundary condition. The coefficient $A$ should be zero due to the constraint as well. Therefore, the electric field and magnetic field in the central region of the rectangular toroidal cavity would not exist when the toroidal cavity has PEC boundary condition.

For non PEC toroidal cavity, which means the cavity surface has a finite conductivity, the electric field could have non zero value at the boundary.  In this case, we assume that the thickness of the boundary goes to zero ($d\ll\delta$) to avoid the decaying effect of the field due to  skin depth, $\delta=\sqrt{\frac{2}{\mu_{c}\omega\sigma}}$.

With this assumption, we can apply continuity condition for both electric and magnetic fields, $\hat{n}\times(\mathbf{E}_{\mathrm{rea}}^{\mathrm{out}}-\mathbf{E}_{\mathrm{rea}}^{\mathrm{in}})=0$, $\hat{n}\times(\mathbf{B}_{\mathrm{rea}}^{\mathrm{out}}-\mathbf{B}_{\mathrm{rea}}^{\mathrm{in}})=0$ at $r=r_{1}$ and $r=r_{2}$ to determine unknown coefficients in the solutions of the electric and magnetic fields in three different regions, $0\leq r<r_{1}$,  $ r_{1}<r<r_{2}$, and $ r_{2}<r$. In $kr\ll 1$ limit, the magnetic field in the central region of the rectangular toroidal cavity where $0\leq r<r_{1}$ becomes
\begin{equation}
\mathbf{B}_{\mathrm{rea}}=ig_{a\gamma \gamma} a \mathrm{B}_{0}kr_{1}\ln (\frac{r_{2}}{r_{1}})\hat{z}=-\frac{g_{A}\dot{ a}\mathrm{B}_{0}r_{1}}{c}\ln (\frac{r_{2}}{r_{1}})\hat{z}.
\end{equation}

\section{\label{sec:level1}Electromagnetic Energy in Cavity Modes}
The total electromagnetic energy stored in the cavity is the sum of each contribution from electromagnetic energy stored in cavity modes as,
\begin{equation}
U_{\mathrm{tot}}=U_{\mathrm{E}}+U_{\mathrm{B}}=\frac{1}{4} \int \left( \epsilon_{0}\mathbf{E}_{\mathrm{rea}}\cdot\mathbf{E}_{\mathrm{rea}}^{*} +\frac{1}{\mu_{0}} \mathbf{B}_{\mathrm{rea}}\cdot\mathbf{B}_{\mathrm{rea}}^{*}\right) dV.
\label{eq4-0-3}
\end{equation}
For the haloscope search with a resonant cavity, it was normally assumed that the electric and magnetic energies stored in the cavity were the same ($U_{\mathrm{E}}=U_{\mathrm{B}}$) for both the cylindrical and toroidal cavity cases as long as $\nabla \times {\mathbf{B}_{0}} = 0$\cite{2016PhRvD..94k1702K,2017PhRvD..96f1102C}. This means that the difference between the electric and magnetic stored energies should be zero,
\begin{equation}
\delta U=U_{\mathrm{E}}-U_{\mathrm{B}}=0.
\label{eq4-0-3}
\end{equation}
However, this has never been analytically verified especially for the toroidal cavity case. We evaluated the stored energy for reacted electric and magnetic fields for both the cylindrical and toroidal cavity cases. In the cylindrical cavity case, the stored energy difference can be estimated from the solution for the electromagnetic field in Eq.\ref{eq3-4} as follows:
\begin{equation}
\begin{aligned}
\delta U&=\frac{1}{4}\int \left(\epsilon_{0}\mathbf{E}_{\mathrm{rea}}\cdot\mathbf{E}^{*}_{\mathrm{rea}}-\frac{1}{\mu_{0}}\mathbf{B}_{\mathrm{rea}}\cdot\mathbf{B}^{*}_{\mathrm{rea}}\right)dV\\
&=-\frac{\pi}{8\mu_{0}} g^{2}_{A} a^{2}\textrm{B}^{2}_{0}r_{0}^{2}L\left(\frac{\mathbf{J}_{2}(kr_{0})}{\mathbf{J}_{0}(kr_{0})}\right).
\end{aligned}
\label{eq4-1}
\end{equation}

As seen in Eq.\ref{eq4-1}, the stored energies for the electric and magnetic fields do not exactly cancel each other out. The energy difference $\delta U$ can be interpreted from the Poynting relationship. Using Eq.\ref{eq2-13}, the following relationship can be derived using the Poynting vector, $\mathbf{S}=\frac{1}{2\mu_{0}}\left(\mathbf{E}\times\mathbf{B^{*}}\right)$, 
\begin{equation}
\frac{\partial u}{\partial t}+\textrm{Re}\left[\nabla \cdot \mathbf{S}\right]=-\frac{1}{2}\mathrm{Re}\left[\mathbf{J}^{*}\cdot \mathbf{E}\right],
\label{eq4-1-1}
\end{equation}
where $u$ is the energy density and $\mathbf{J}$ is a current. Since the boundary condition for haloscope searches already requires a zero normal current ($\mathbf{J_{e}}=0$), $\mathbf{J}$ should not be induced from a normal electromagnetic interaction but could be induced from the axion-photon interaction. Therefore, it can be interpreted as non-dissipating axion induced current as $\mathbf{J}=\mathbf{J}_{\textrm{a}}$. Using the similarity from the relationship between polarization density and polarization density current as $\mathbf{J}_{p}=\partial\mathbf{P}/\partial t$, one can set 
\begin{equation}
\mathbf{J}_{a}=\frac{\partial\mathbf{P}_{a}}{\partial t}=-\sqrt{\frac{\epsilon_{0}}{\mu_{0}}}g_{a\gamma\gamma}\frac{\partial a}{\partial t}\mathbf{B}_{0},
\end{equation}
where $\mathbf{P}_{a}=-\sqrt{\frac{\epsilon_{0}}{\mu_{0}}}g_{A} a\mathbf{B}_{0}$ is an axion induced polarization density with a unit of $\textrm{A}\cdot s/ m^{2}$ and $\mathbf{J}_{\textrm{a}}$ oscillates with axion frequency $\omega_{a}$.

The energy difference $\delta U$ comes from the axion induced polarization density $\mathbf{P}_{a}$ as 
\begin{equation}
\delta U=-\frac{1}{4}\int \mathrm{Re} \left[ \mathbf{P}_{a}\cdot\mathbf{E}^{*}_{\mathrm{rea}} \right] dV.
\end{equation}

\begin{figure}
\centering
\begin{subfigure}[H]{0.4\textwidth}
\includegraphics[width=\textwidth]{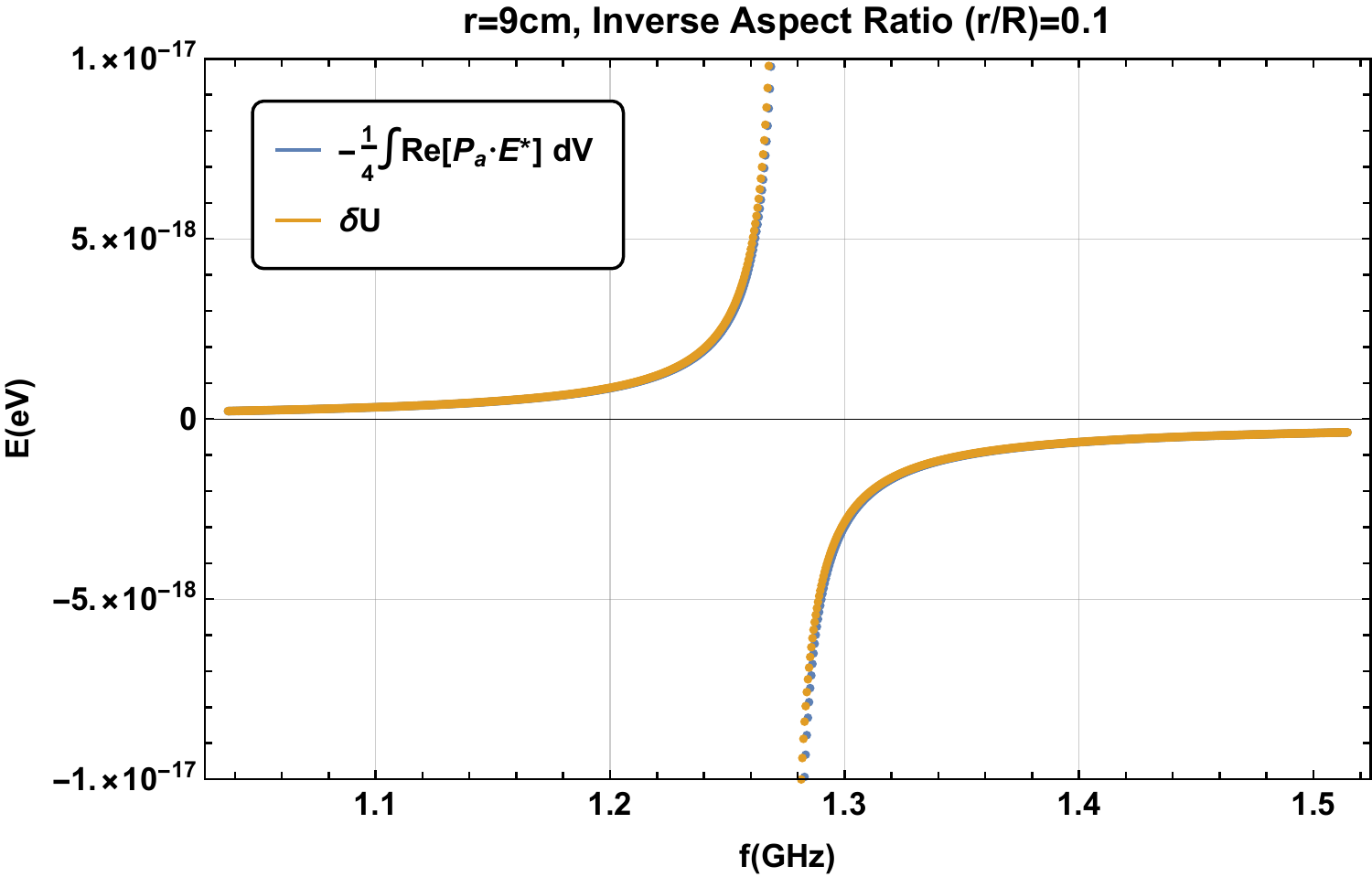}
\caption{}
\end{subfigure}
\begin{subfigure}[H]{0.4\textwidth}
\includegraphics[width=\textwidth]{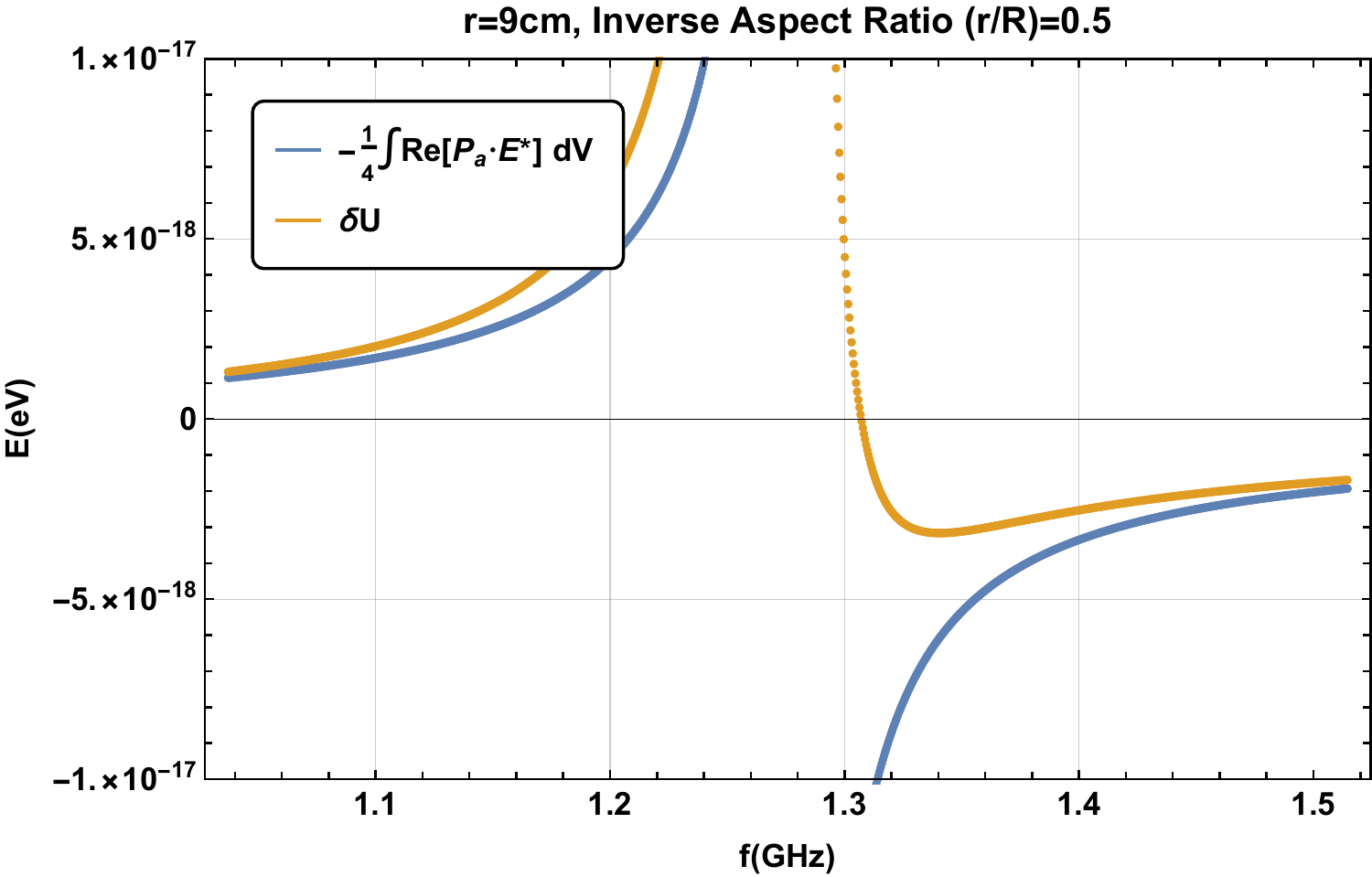}
\caption{}
\end{subfigure}
\caption{The energy difference as a function of resonant frequency in a toroidal cavity with different inverse aspect ratio: (a) $r_{0}/R=0.1$, (b) $r_{0}/R=0.5$}
\label{fig1}
\end{figure}

In the toroidal cavity case, the difference between the electric and magnetic stored energies can be estimated from the solutions in Eq.\ref{eq3-7}. The only difference between $\delta U$ and $-\frac{1}{4}\int \mathrm{Re} \left[ \mathbf{P}_{a}\cdot\mathbf{E}^{*}_{r}  \right] dV$ is the integration part for $r$ mainly due to the approximation as shown in Eq.\ref{eq37-a} and Eq.\ref{eq37-b} respectively,
\begin{subequations}
\begin{align}
&\int_{0}^{r_{0}}-2r\frac{(\mathbf{J}_{0}(kr_{0})-\mathbf{J}_{0}(kr))^{2}-\mathbf{J}_{1}(kr)^{2}}{\sqrt{R^{2}-r^{2}}\mathbf{J}_{0}(kr_{0})^{2}}dr,\label{eq37-a}\\
&\int_{0}^{r_{0}}-2r\frac{\mathbf{J}_{0}(kr_{0})-\mathbf{J}_{0}(kr)}{\sqrt{R^{2}-r^{2}}\mathbf{J}_{0}(kr_{0})}dr\label{eq37-b}.
\end{align}
\label{eq37}
\end{subequations}
The difference was numerically estimated with the conditions of a minor radius $r_{0}=9~\mathrm{cm}$, and the inverse aspect ratio $\eta=0.1,\ 0.5$ respectively as shown in Fig \ref{fig1}.

The ratio of the energy difference over total energy, 
\begin{equation}
\mathcal{R}=\left| \frac{ \delta U}{U_{\textrm{tot}}} \right|, 
\end{equation}
was estimated for a cylindrical cavity as well as a toroidal cavity. The geometry for each case is shown in Table \ref{table1}. For the cylindrical cavity, the $\mathcal{R}$ was estimated for a geometry with a radius $r=9\ \mathrm{cm},$ and height $L=50\ \mathrm{cm}$, as a function of resonant frequency $f$ shown in Fig. \ref{fig1}. The difference in resonance is lower than $10^{-4}$ but gets larger, up to $10^{-1}$ as the frequency moves toward the off resonant regime. The geometry of the toroidal cavity was taken from ACTION\cite{2017PhRvD..96f1102C}. The difference is almost identical to the cylindrical case but the resonant frequencies are slightly different between the two inverse aspect ratios $\epsilon=0.1,\ 0.5$. 
\begin{figure}[h]
\centering
\includegraphics[width=0.4\textwidth]{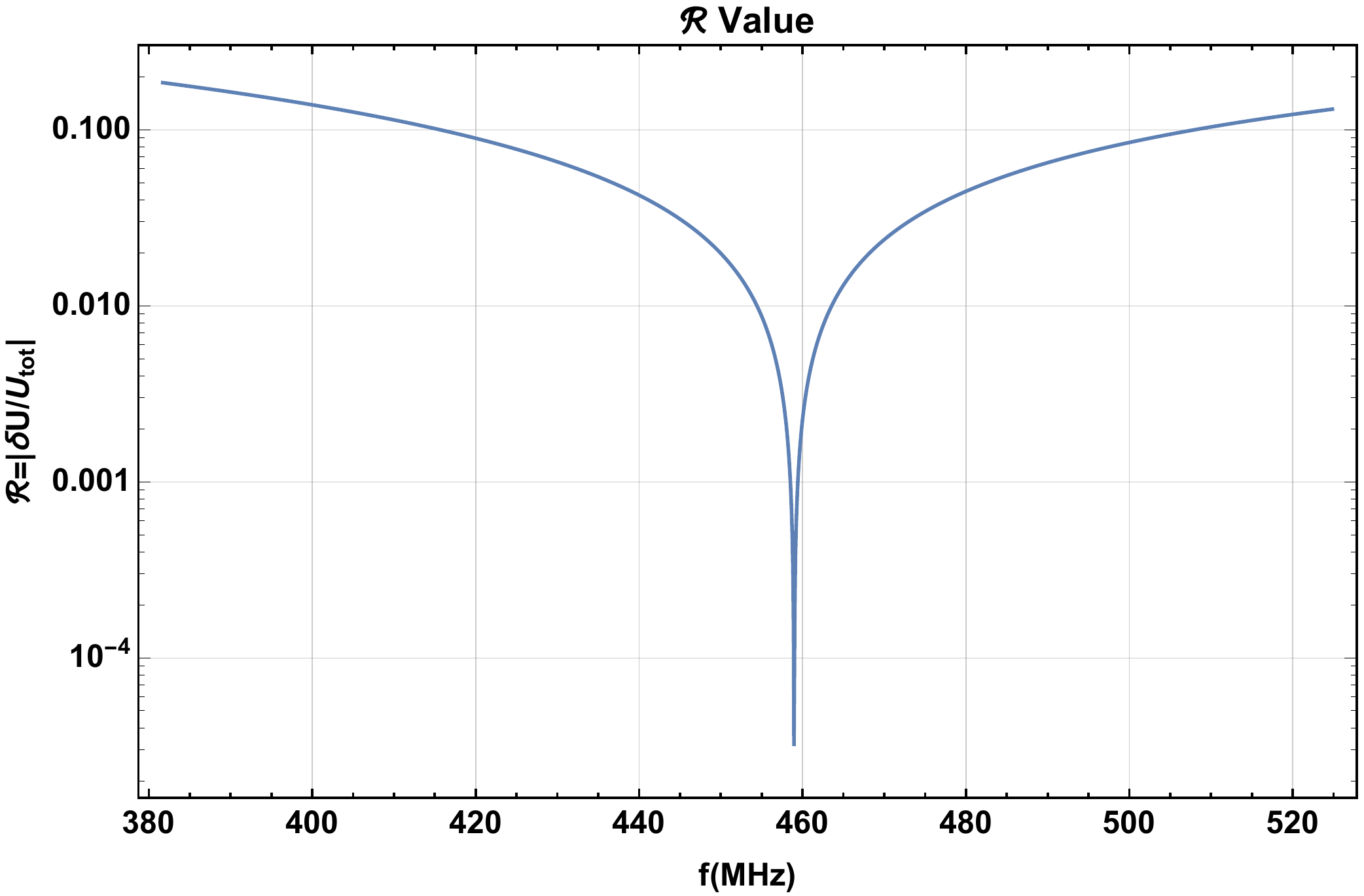}
\caption{The ratio of energy difference as a function of resonant frequency $f$ in a cylindrical cavity}
\label{fig2}
\end{figure}

\begin{figure}[h]
\centering
\begin{subfigure}{0.4\textwidth}
\includegraphics[width=\textwidth]{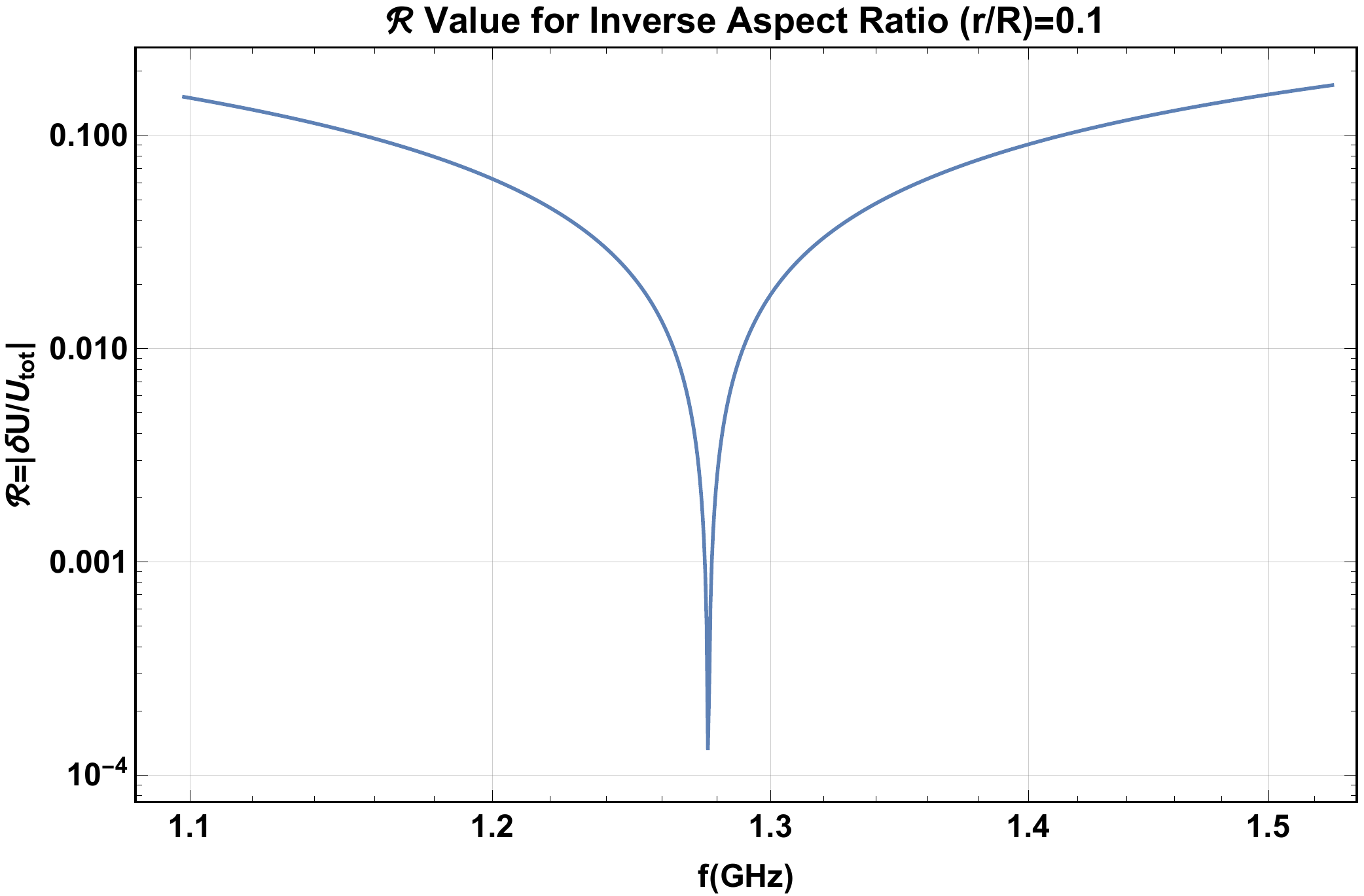}
\caption{}
\end{subfigure}
\begin{subfigure}{0.4\textwidth}
\includegraphics[width=\textwidth]{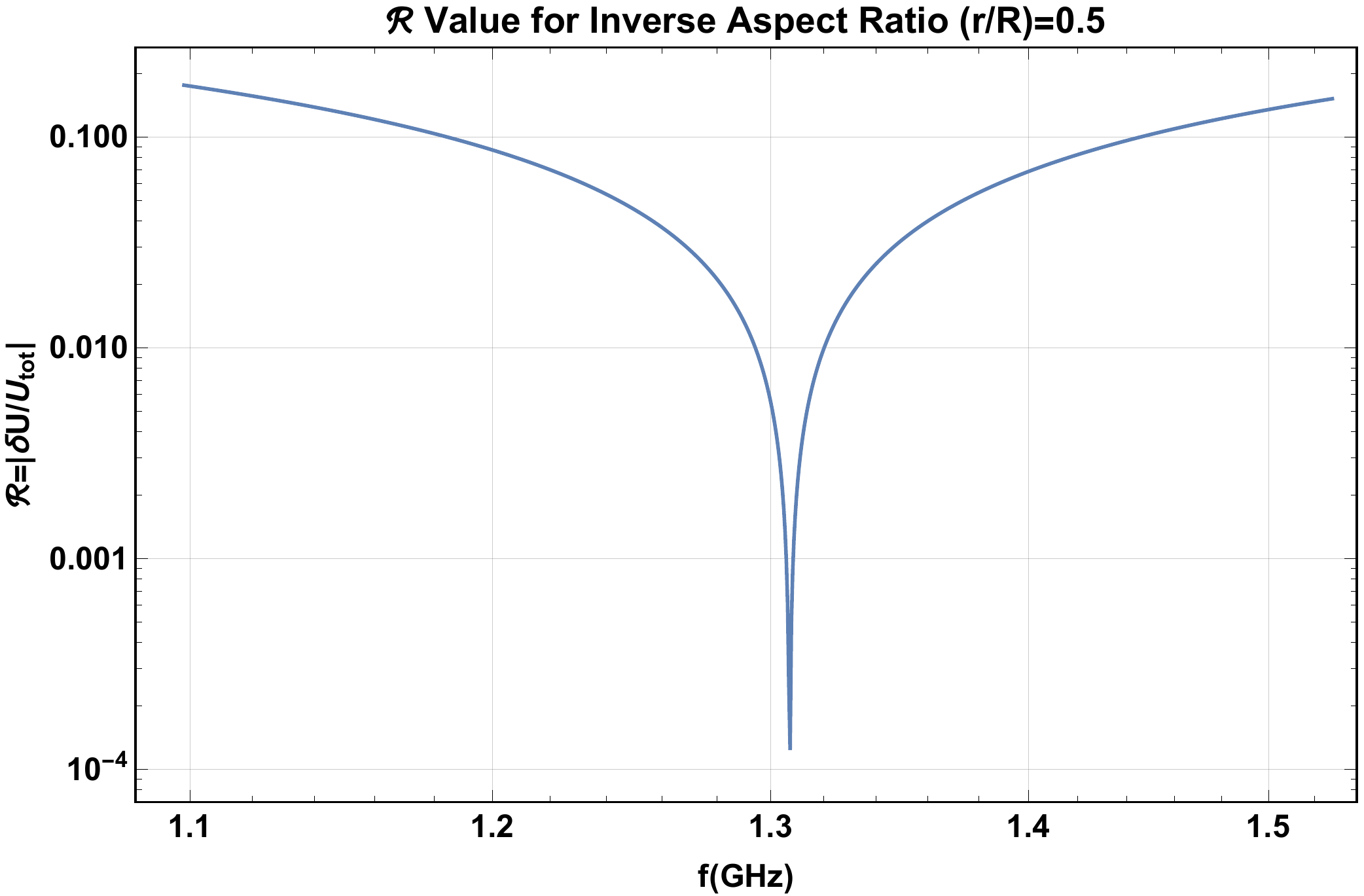}
\caption{}
\end{subfigure}
\caption{The ratio of energy difference as a function of resonant frequency in a toroidal cavity with different inverse aspect ratios: (a) $r_{0}/R=0.1$, (b) $r_{0}/R=0.5$}
\label{fig3}
\end{figure}

\begin{table}
\begin{center}
\begin{tabular}{c c c c c c}
\hline
 &Cylinder& & Toroid \\
\hline
r & 25cm&r&9cm\\
\hline
L &100cm &R&50cm\\
\hline
f & 458.9MHz &f & 1.27GHz\\
\hline
$m_{a}$&1.90$\mu$eV&$m_{a}$&5.27$\mu$eV\\
\hline
\end{tabular}
\end{center}
\caption{The geometries of the cylindrical and toroidal cavities. $f$ is the resonant frequency in the lowest mode and $m_{a}$ is the corresponding mass of the axion.}
\label{table1}
\end{table}
\section{\label{sec:level1}Conclusion}
We have applied an effective approximation into Maxwell's equations describing axion-photon coupling to acquire Mexwell's equations only for the reacted electromagnetic field for haloscope searches. We evaluated the resonant electromagnetic field for two different resonant cavity geometries, a cylindrical cavity and a toroidal cavity, from the decoupled Maxwell's equations for reacted electromagnetic fields. We also showed that the electromagnetic field can be approximated from the resonant solutions. Stored energy was evaluated for both cavity cases. A difference arises between the electric and magnetic energies stored in the cavity mode. The difference can be interpreted as a polarization density induced by the axion photon interaction.
\section{\label{sec:level1}Acknowledgement}
This work was supported by IBS-R017-D1-2019-a00. 
%
\appendix 

\section{APPENDIX : Integration form of Faraday's Law}
One can also check whether the electromagnetic field satisfies the Faraday's law in Eq.\ref{eq2-7-1-2} with an integration form as follows:
\begin{equation}
\int \nabla \times\mathbf{E}_{a} \cdot d\mathbf{A}=-\frac{d}{dt}\int \mathbf{B}_{a}\cdot d\mathbf{A}.
\label{appeq-002}
\end{equation}
Using Stoke's theorem, Eq.\ref{appeq-002} becomes,
\begin{equation}
\oint \mathbf{E}_{a} \cdot d\mathbf{l}=-\frac{d}{dt}\int \mathbf{B}_{a}\cdot d\mathbf{A}.
\label{appeq-001}
\end{equation}
\subsection{A. Approach from Ref.\cite{2016PhRvL.116p1804M, 2016PhRvD..94k1702K}}
From the solution for electric field and magnetic field in Eq.\ref{eq2-7-5} and Eq.\ref{eq2-7-3}, $\mathbf{E}_{a}$, and $\mathbf{B}_{a}$ become 
\begin{subequations}
\begin{align}
\mathbf{E}_{a}&=-cg_{a\gamma \gamma} a \mathrm{B}_{0}\hat{z},\\
\mathbf{B}_{a}&=-g_{a\gamma \gamma}\mathrm{B}_{0}\dot{ a}/2c\hat{\phi}.
\end{align}
\end{subequations}
One can check the relationship in Eq.\ref{appeq-001} from the electromagnetic field. The integration area can be arbitrary. But for simplicity, one can assume a rectangular area with either symmetric or asymmetric cases.
\subsubsection{a. Integration for symmetric rectangular area}
If the integration area is bound by a symmetric rectangle (with corners $a,\ b,\ c,\ d$) with respect to the symmetric axis of a solenoid,
\begin{figure}[H]
\begin{center}
\includegraphics[width=0.7\linewidth]{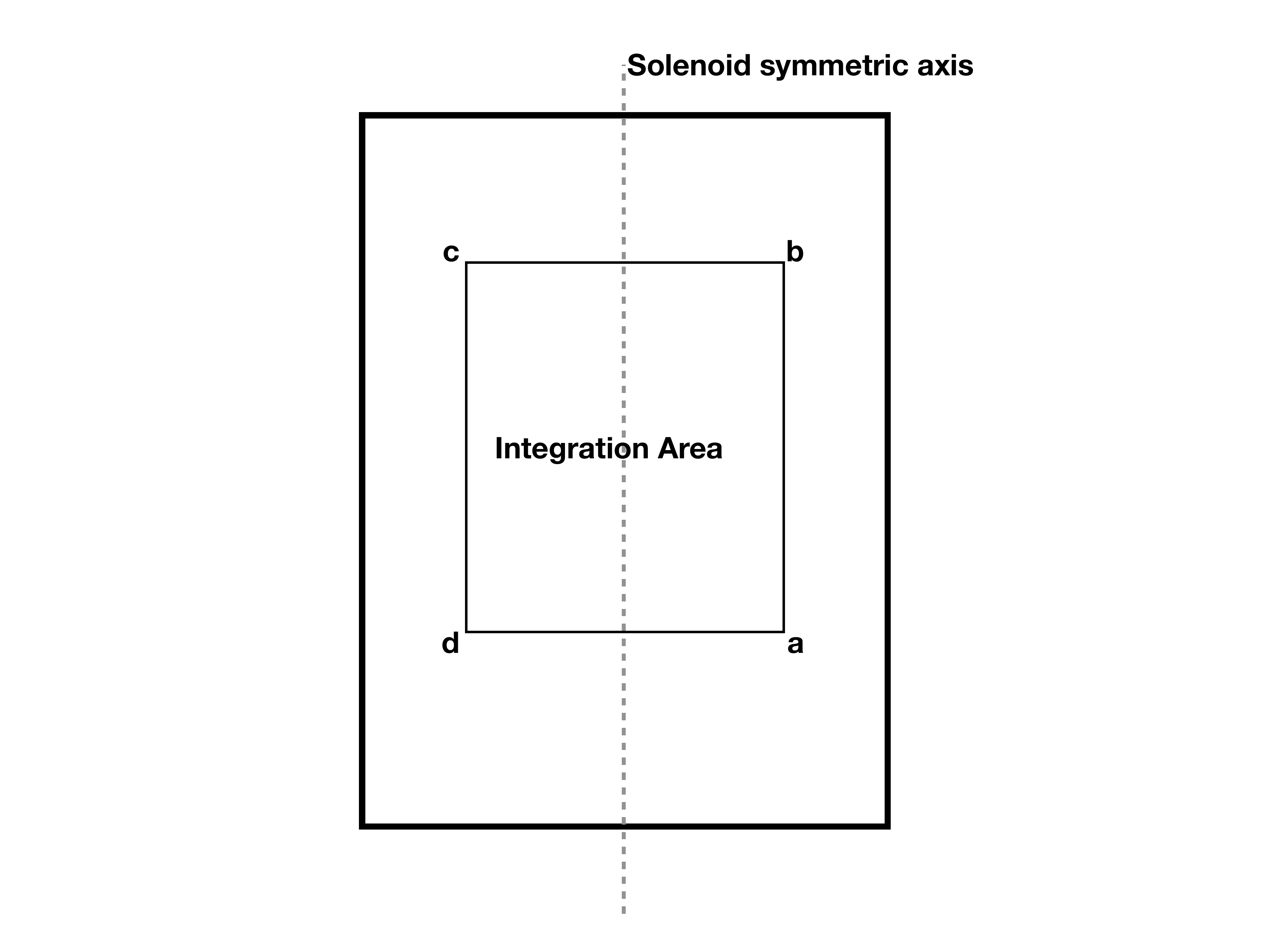}
\end{center}
\caption{Integration area for symmetric axis}
\end{figure}
The line integration of the electric field with a symmetric property becomes,
\begin{equation}
\begin{aligned}
\oint \mathbf{E}_{a} \cdot d\mathbf{l}&=\int_{a}^{b}\mathbf{E}_{a}dl+\int_{b}^{c}\mathbf{E}_{a}dl+\int_{c}^{d}\mathbf{E}_{a}dl+\int_{d}^{a}E_{a}dl\\
&=\int_{a}^{b}E_{a}dl-\int_{d}^{c}E_{a}dl=0,
\end{aligned}
\end{equation}
and the surface integration of the magnetic field becomes,
\begin{equation}
\begin{aligned}
\int \mathbf{B}_{a}\cdot d\mathbf{A}&=\int \mathbf{B}_{a} \cdot d\mathbf{A}_{1}+\int \mathbf{B}_{a} \cdot d\mathbf{A}_{2}\\
&=\int \mathrm{B}_{a}dA_{1}-\int \mathrm{B}_{a} dA_{2}=0.
\end{aligned}
\end{equation}
where the normal vector of $d\mathbf{A}$ is $\hat{n}$.The integration area is divided into two sections, one is the right side of the symmetric axis, and the other is the left side of the symmetric axis. Then, because one of the magnetic field directions is parallel to the $\hat{n}$, and the other is anti parallel, the surface integration becomes zero. Therefore, with this integration along the symmetric area, those solutions for electric and magnetic fields still satisfy Faraday's law.
\subsubsection{b. Integration area for asymmetric rectangle}
Let's consider the asymmetric rectangle (with corners $a,\ b,\ c,\ d$) where the edge $\overline{da}$ is aligned to the symmetric axis.
\begin{figure}[H]
\begin{center}
\includegraphics[width=0.7\linewidth]{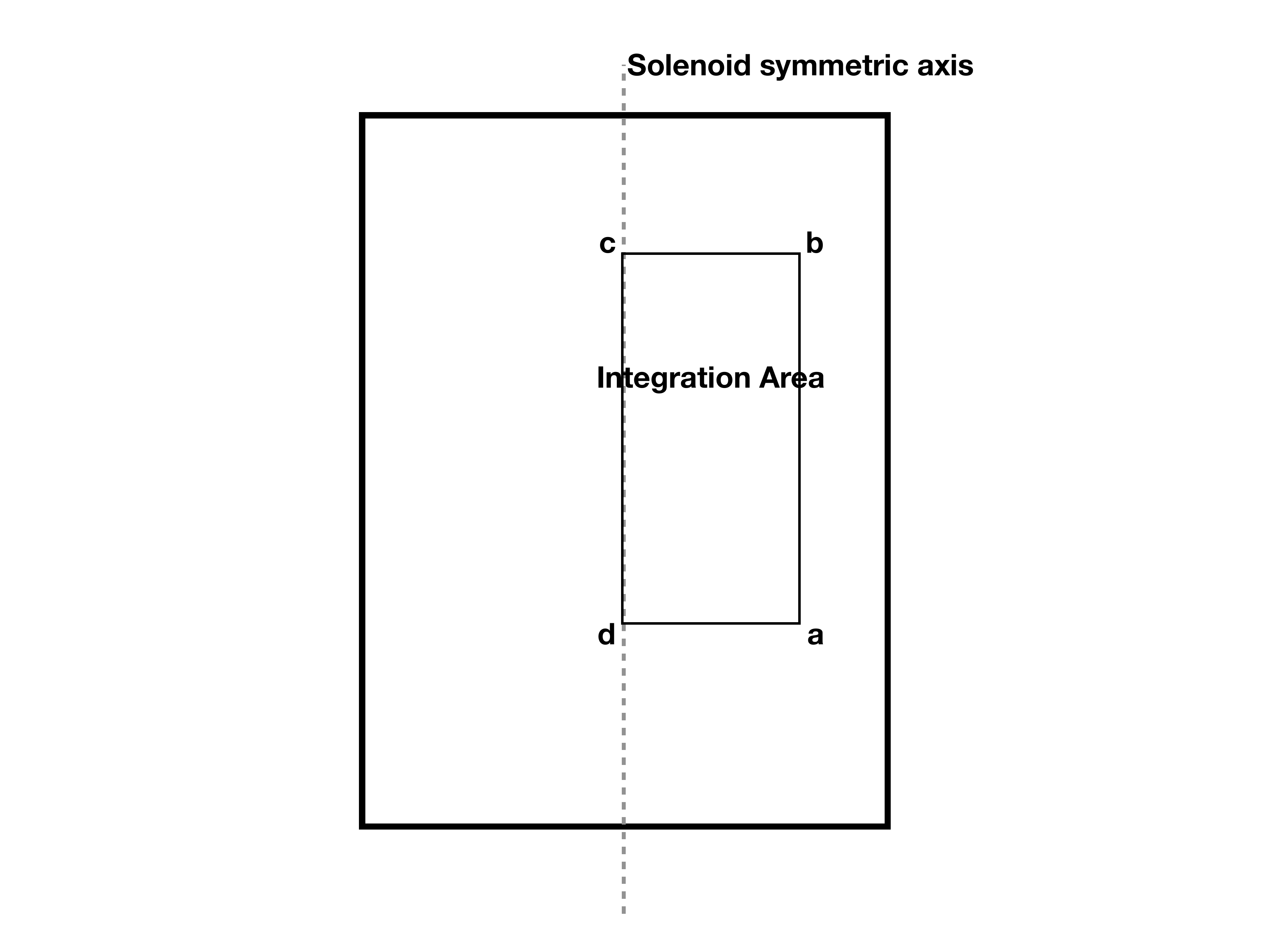}
\end{center}
\caption{Integration area for asymmetric axis}
\end{figure}
 The line integration becomes zero because the magnetic field is uniform over the space. Even if we consider $\mathbf{f}(r)$ from Eq.\ref{eq2-7-5}, the left line integration part becomes,
\begin{equation}
\oint \mathbf{E}_{a} \cdot d\mathbf{l}=\oint \mathbf{f}(r)d\mathbf{l}.
\end{equation}
At the same time,  the surface integration term becomes, 
\begin{equation}
\int\mathbf{B}_{a}\cdot d\mathbf{A}=-\frac{1}{2c}g_{a\gamma \gamma}\mathrm{B}_{0}\dot{ a}\hat{\phi}\cdot{n}\int_{0}^{L}\int_{0}^{R} rdr=-\frac{g_{a\gamma \gamma}\mathrm{B}_{0}\dot{ a}}{4c}LR^{2},
\end{equation}
which is not only non zero but also time dependent. The time derivative gives,
\begin{equation}
-\frac{d}{dt}\int \mathbf{B}_{a}\cdot d\mathbf{A}=\frac{g_{a\gamma \gamma}\mathrm{B}_{0}\ddot{ a}}{4c}LR^{2}\neq \oint \mathbf{f}(r)d \mathbf{l}.
\label{appeq-0-1}
\end{equation}
Each side of Eq.\ref{appeq-0-1} can not be equal because $\mathbf{f}$ is not time dependent. The only way to satisfying Eq.\ref{appeq-0-1} is if the double time derivative of the axion field is constant, $\ddot{ a}=\alpha$, which is contradictory to the oscillating axion field. Therefore the Faraday's law is not satisfied with the asymmetric integration surface, and using this solution, $\mathbf{E}_{a}=-cg_{a\gamma \gamma} a \mathrm{B}_{0}\hat{z},\ \mathbf{B}_{a}=-g_{a\gamma \gamma}r\mathrm{B}_{0}\dot{ a}/2c\hat{\phi}$, the Faraday's law is also not generally satisfied in integration form.
\subsection{B. Effective approximation}
In effective approximation, the special solutions for electric field and magnetic field are $\mathbf{E}_{\mathrm{rea}}^{\mathrm{special}}=cg_{a\gamma \gamma} a \mathrm{B}_{0}\hat{z}, \mathbf{B}_{\mathrm{rea}}^{\mathrm{special}}=0$. If one takes the line integration for the electric field, it becomes,
\begin{equation}
\begin{aligned}
\oint \mathbf{E}_{\mathrm{rea}}^{\mathrm{special}} \cdot d\mathbf{l}&=\int_{a}^{b}+\int_{b}^{c}+\int_{c}^{d}\mathbf{E}_{\mathrm{rea}}^{\mathrm{special}}dl+\int_{d}^{a}\mathbf{E}_{\mathrm{rea}}^{\mathrm{special}}dl\\
&=\int_{a}^{b} \mathbf{E}_{\mathrm{rea}}^{\mathrm{special}} dl-\int_{d}^{c} \mathbf{E}_{\mathrm{rea}}^{\mathrm{special}} dl=0.
\end{aligned}
\end{equation}
Since there is no magnetic field corresponding to $\mathbf{E}_{\mathrm{rea}}^{\mathrm{special}}$, the Faraday's law is therefore naturally satisfied.

\subsection{C. Effective approximation in the cylindrical cavity case}
\begin{figure}[H]
\begin{center}
\includegraphics[width=0.7\linewidth]{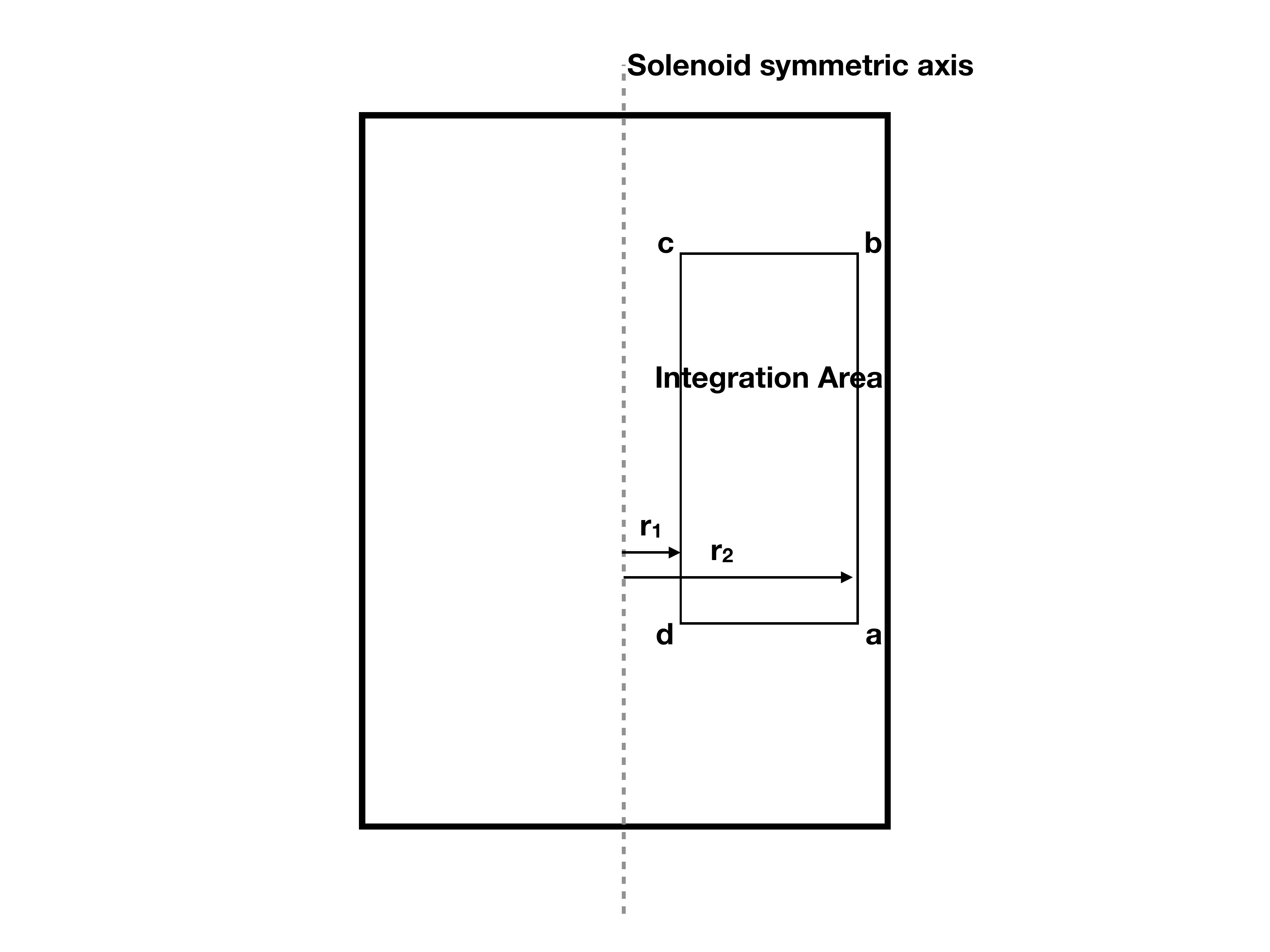}
\end{center}
\caption{Integration area for asymmetric axis}
\label{appfig-01}
\end{figure}

From the effective approximation solution Eq.\ref{eq3-4}, the solution for a cylindrical cavity is,
\begin{equation}
\begin{aligned}
&\mathbf{E}_{\mathrm{rea}}=c g_{a\gamma\gamma} a_{0}\textrm{B}_{0}e^{-i\omega_{a}t}\left(1-\frac{\mathbf{J}_{0}(kr)}{\mathbf{J}_{0}(kr_{0})}\right)\hat{z},\\
&\mathbf{B}_{\mathrm{rea}}=ig_{a\gamma\gamma} a_{0}\textrm{B}_{0}e^{-i\omega_{a}t}\left(\frac{\mathbf{J}_{1}(kr)}{\mathbf{J}_{0}(kr_{0})}\right)\hat{\phi}.
\end{aligned}
\end{equation}
The electric field solution is separated into a special solution $\mathbf{E}_{{s}}$ part and an ordinary $\mathbf{E}_{{n}}$ part as follows:
\begin{equation}
\begin{aligned}
&\mathbf{E}_{{s}}=cg_{a\gamma\gamma} a_{0}\textrm{B}_{0}e^{-i\omega_{a}t},\\
&\mathbf{E}_{{n}}=-cg_{a\gamma\gamma} a_{0}\textrm{B}_{0}\frac{\mathbf{J}_{0}(kr)}{\mathbf{J}_{0}(kr_{0})}e^{-i\omega_{a}t}.
\end{aligned}
\end{equation}
Since it was shown that the special solution satisfies Faraday's law in the previous section, now one can check the relation between the ordinary solution for electric field $\mathbf{E}_{{n}}$ and magnetic field $\mathbf{B}_{{r}}$. If one takes a line integral for the ordinary electric field as shown in Fig.\ref{appfig-01},
\begin{equation}
\oint \mathbf{E}_{n}\cdot d\mathbf{l}=\int_{a}^{b}\mathbf{E}_{n}dl+\int_{b}^{c}\mathbf{E}_{n}dl+\int_{c}^{d}\mathbf{E}_{n}dl+\int_{d}^{a}\mathbf{E}_{n}dl.
\end{equation}
Since the direction $\overline{da},\ \overline{bc}$ is perpendicular to the electric field, the integration path is reduced by $\overline{ab},\ \overline{cd}$. Since these integrations are constant over the integration $dl$, the integration becomes,
\begin{equation}
\oint \mathbf{E}_{n}\cdot d\mathbf{l}=-\frac{cg_{a\gamma \gamma} a_{0} \mathrm{B}_{0}L}{\mathbf{J}_{0}(kr_{0})}\left(\mathbf{J}_{0}(kr_{2})-\mathbf{J}_{0}(kr_{1})\right)e^{-i\omega_{a}t},
\label{appeq-04}
\end{equation}
here the radius has inequality, $0<r_{1}<r_{2}<r_{0}$. The remaining integration is surface integration over the magnetic field,
\begin{equation}
\begin{aligned}
&-\frac{d}{dt}\int \mathbf{B}_{\mathrm{rea}}\cdot d\mathbf{A}\\
&=-\omega_{a}g_{a\gamma \gamma} a_{0}\mathrm{B}_{0}L\int_{r_{1}}^{r_{2}}dr \frac{\mathbf{J}_{1}(kr)}{\mathbf{J}_{0}(kr_{0})}e^{-i\omega_{a}t}\hat{\phi}\cdot\hat{n}\\
&=-\omega_{a}g_{a\gamma \gamma} a_{0}\mathrm{B}_{0}L\left(\frac{\mathbf{J}_{0}(kr_{1})-\mathbf{J}_{0}(kr_{2})}{k\mathbf{J}_{0}(kr_{0})}\right)e^{-i\omega_{a}t}\hat{\phi}\cdot\hat{n}\\
&=cg_{a\gamma \gamma} a_{0}\mathrm{B}_{0}L\left(\frac{\mathbf{J}_{0}(kr_{1})-\mathbf{J}_{0}(kr_{2})}{\mathbf{J}_{0}(kr_{0})}\right)e^{-i\omega_{a}t},\\
\end{aligned}
\label{appeq-05}
\end{equation}
By comparing the results from Eq.\ref{appeq-04} and Eq.\ref{appeq-05}, one can see the solutions from the effective approximation.

When the integration area is crossed over the symmetric axis, the $\hat{\phi}$ is flipped with respect to the symmetric axis as shown in Fig.\ref{appfig-02}. In this case, the electric field is straightforward as, 
\begin{equation}
\oint \mathbf{E}_{n}\cdot d\mathbf{l}=-{cg_{a\gamma \gamma} a_{0} \mathrm{B}_{0}L}\left(\frac{\mathbf{J}_{0}(kr_{2})-\mathbf{J}_{0}(kr_{1})}{\mathbf{J}_{0}(kr_{0})}\right)e^{-i\omega_{a}t}.
\end{equation}
However, for the magnetic field calculation, it is necessary to separate the integration into two parts, one parallel to the $\hat{n}$, and anti parallel to the $\hat{n}$, as follows:
\begin{figure}[H]
\begin{center}
\includegraphics[width=0.7\linewidth]{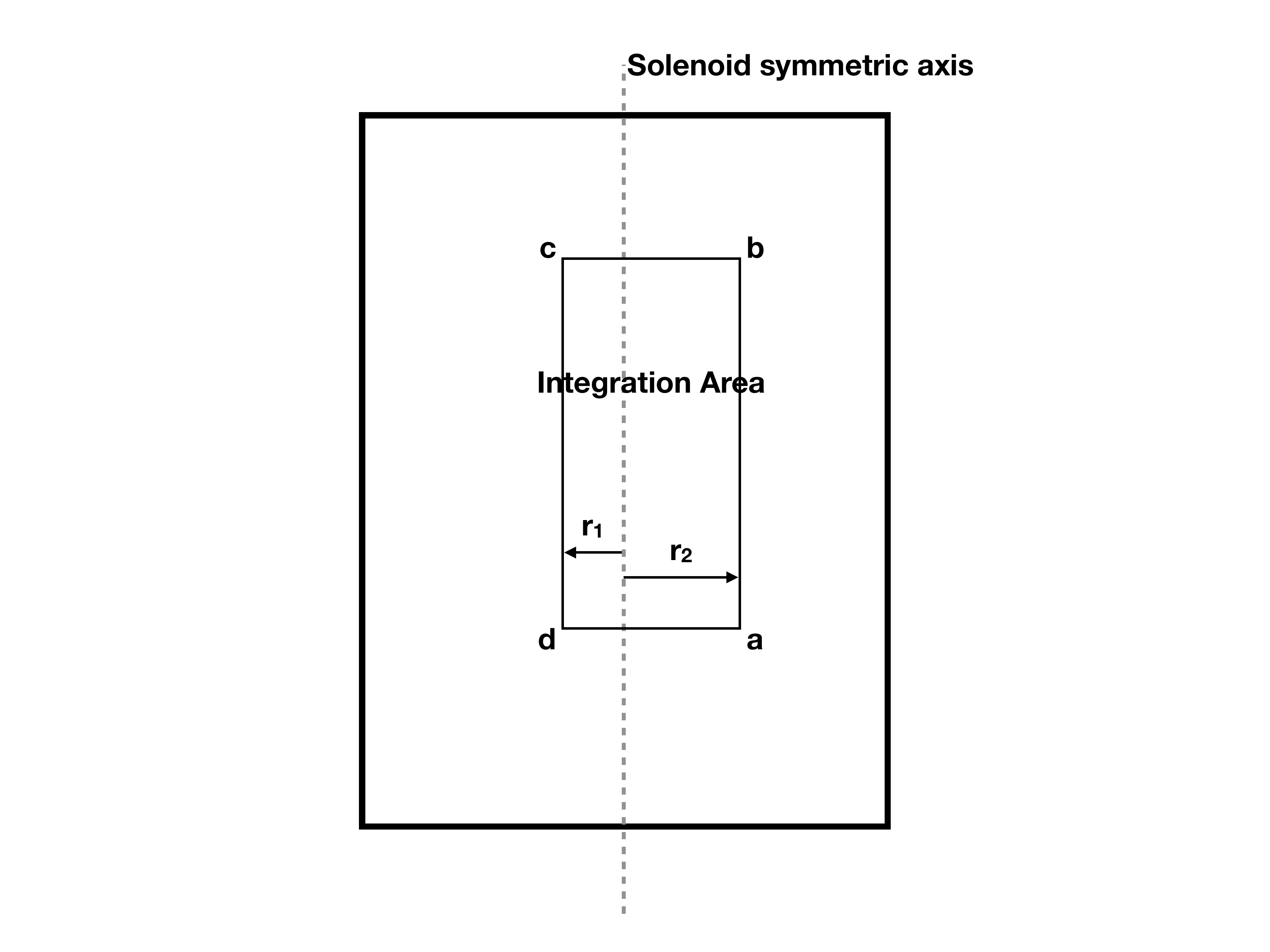}
\end{center}
\caption{Integration area for asymmetric axis}
\label{appfig-02}
\end{figure}
\begin{equation}
\begin{aligned}
&-\frac{d}{dt}\int \mathbf{B}_{\mathrm{rea}}\cdot d\mathbf{A}=-\frac{d}{dt}\left(\int B_{r}dA_{1}-\int B_{r}dA_{2}\right),\\
\end{aligned}
\end{equation}
here we set the $dA_{1}$ as parallel and $dA_{2}$ is anti-parallel case.
\begin{equation}
\begin{aligned}
&\int B_{r}dA_{1}-\int B_{r}dA_{2},\\
&=L(\int_{0}^{r_{1}}B_{r}dr-\int_{0}^{r_{2}}B_{r}dr),\\
&=\frac{ig_{a\gamma \gamma} a_{0} \mathrm{B}_{0}}{k\mathbf{J}_{0}(kr_{0})}((\mathbf{J}_{0}(0)-\mathbf{J}_{0}(kr_{1}))-(\mathbf{J}_{0}(0)-\mathbf{J}_{0}(kr_{2}))),\\
&=\frac{ig_{a\gamma \gamma} a_{0} \mathrm{B}_{0}}{k\mathbf{J}_{0}(kr_{0})}(\mathbf{J}_{0}(kr_{2})-\mathbf{J}_{0}(kr_{1})),\\
\end{aligned}
\end{equation}
and do the full calculation, we get,
\begin{equation}
-\frac{d}{dt}\int \mathbf{B}_{\mathrm{rea}}\cdot d\mathbf{A}=-{cg_{a\gamma \gamma} a_{0} \mathrm{B}_{0}L}\left(\frac{\mathbf{J}_{0}(kr_{2})-\mathbf{J}_{0}(kr_{1})}{\mathbf{J}_{0}(kr_{0})}\right)e^{-i\omega_{a}t},
\end{equation}
which is the same as the line integration of the electric field. Therefore, for any rectangle integration area, the solutions for the electric and magnetic fields from the effective approximation still fully satisfy the Faraday's law.
\end{document}